\documentclass[preprint]{aastex}  

\begin{document}                                                                     

\title{High Mass Triple Systems: The Classical Cepheid Y Car           
  \altaffilmark{1} }


\author{Nancy Remage Evans }
\affil{Smithsonian Astropnysical Observatory, Cambridge, MA 02138}

\author{Kenneth G. Carpenter}
\affil{Goddard Space Flight Center}

\author{Richard Robinson}
\affil{Johns Hopkins University}

\author{Francesco Kienzle}
\affil{Geneva Observatory}


\author{Anne E. Dekas}
\affil{Harvard College}

\email{nevans@cfa.harvard.edu, kgc@stargate.gsfc.nasa.gov,robinson@pha.jhu.edu,
Francesco.Kienzle@obs.unige.ch}


\altaffiltext{1}{Based on observations made with the NASA/ESA Hubble Space                
             Telescope, obtained at the Space Telescope Science                       
             Institute, which is operated by the Association of                       
             Universities for Research in Astronomy, Inc. }


\begin{abstract}

We have obtained an HST STIS ultraviolet high dispersion Echelle mode spectrum 
the binary companion 
of the double mode classical Cepheid Y Car.  The 
velocity measured for the hot companion from this 
spectrum is very different from reasonable predictions for binary motion, 
implying that the companion is itself a short period
binary.  The measured velocity  changed by 
7 km sec$^{-1}$ during the 4 days between two 
segments of the observation confirming  this 
interpretation.  We  summarize ``binary" 
Cepheids which are in fact members of triple 
system and find at least 44\% are triples. 
The summary of information on Cepheids with orbits 
makes it likely that the fraction is under-estimated.

\end{abstract}




\section{Introduction}

Because of the importance of classical Cepheid variable
stars as primary extragalactic 
distance indicators, it is extremely important to 
understand them as fully as possible.  Comparison with 
evolutionary tracks is an important test of this 
understanding.  Conversely, because Cepheids have  a 
well defined luminosity from the period--luminosity--color
relation (PLC) they provide a uniquely strong test for 
the evolutionary tracks of objects evolved beyond the 
main sequence.  The inconsistency between 
the masses predicted from evolutionary tracks and 
pulsation calculations (called the ``Cepheid mass 
problem" since the 1960's) has now been resolved 
with a reevaluation of envelope opacities  
(Iglesias and Rogers  1991).  On the other hand, 
the treatment of the boundary between the convective
core and the radiative envelope in intermediate mass 
main sequence stars (``core overshooting) is  
a significant uncertainty in computing evolutionary tracks which 
go through the Cepheid instability strip.   

Linking  a measured mass for a Cepheid  
with its precise luminosity provides a very strong test
of evolutionary calculations.  The advent of satellite 
ultraviolet spectroscopy has made it possible for the first 
time to measure  the orbital velocity amplitude of 
a few hot companions of Cepheids. The mass of the Cepheid can 
then be obtained by combining this amplitude
with the orbital velocity amplitude of the Cepheid from the 
ground-based orbit and a mass inferred from the spectral type
of the companion.
A summary of these measured Cepheid masses provided by 
IUE and HST spectroscopy is given in Evans et al. (1998).

Double mode Cepheids have been particularly important in 
discussions of observed and theoretical masses.  In addition 
to masses inferred from a ``pulsation constant" (theoretical
period-mass-radius relation), a mass can be determined 
from theoretical predictions of the period ratios of the 
fundamental and first overtone periods.  Y Car 
= HD 91595 is a bright 
double mode Cepheid  that was discovered by Stobie and Balona
(1979) to be a member of a binary system.  An orbit 
was derived by Balona (1983).  The hot companion to the Cepheid
was detected by Evans (1992a) in an IUE observation 
 from which the luminosity of the Cepheid was 
determined together with the spectral type of the companion.
A study with the HST GHRS (B\"ohm--Vitense et al. 1997) 
obtained velocities
of the companion and produced a measurement of the mass 
although with large error bars of $\pm$ 32\%.

\section{Observations}

In order to improve the accuracy of the velocities of the companion
Y Car B, we observed it using the Hubble Space Telescope (HST)
Space Telescope  Imaging Spectrograph (STIS) in the Echelle high 
dispersion  mode, E230H, which provides a resolution of
$\lambda$/228000 (Kimble, et al. 1998). 
The wavelength range was centered on 1763 \AA, 
with a typical dispersion of  1.3 km sec$^{-1}$px$^{-1}$
The observation was taken over 8 orbits, 5 on May 3, 2002, 3 on 
May 7, 2002.  Several aspects of the observation were designed to 
obtain the maximum velocity accuracy.  Special wavelength calibration
observations were obtained at the beginning and end of each orbit.
The exposure time was 10 times longer than the standard 
WAVCAL observations in order to bring up faint lines, and provide 
more counts in stronger lines.  In order for the target to be 
centered in the aperture as precisely as possible, a ``peakup"
was done at the beginning of the first and fourth orbits on the 
first day, and the first orbit on the second day.   
As a result of the wavelength calibrations and the target centering,                 
the line centers are expected to be located to an accuracy of                        
0.2 px or 0.26 km sec$^{-1}$.                                                        

In addition to observations of Y Car, we also obtained data on the A1 V
standard star HD 72660. This star has a spectral type near that of the Cepheid
companion and  a very low rotation velocity. 
It's  velocity (Fekel, 1996, using 68 Tau as a comparison) is 2.9
$\pm$ 0.2 km/s. 

The observations for both stars were analyzed using the
CALSTIS software, which was developed for the STIS IDT 
(Instrument Definition Team) at Goddard Space Flight
Center. This software applied a flat field to the data and removed the effects
of bad pixels, hot pixels and Doppler motions. It then located each Echelle
order on the detector and determined the background intensity by measuring
between the echelle orders. The individual orders were then extracted,
background subtracted and  a wavelength and flux calibration
were applied. The 8
individual observations were then combined into a final spectrum.
                                                                 
To determine the velocity of the Cepheid companion  we cross-correlated each of the  
extracted orders of the final Y Car observation with the corresponding order        
in the HD 72660 observation, after placing both on the same wavelength scale.        
Each correlation then corresponds to an independent estimate of the relative velocity.
The correlation functions were examined and it was found that 29 of the first 32      
orders (covering the wavelength region between 1760 and 1900 \AA) had well defined peaks 
in their correlations. For higher orders the correlations disintegrated and gave no   
useful results. This was  caused by the decreasing S/N resulting from        
the rapid decrease in spectrograph sensitivity towards shorter wavelengths.             
The velocities measured from the 29 reliable orders range from -35 to -47 km s$^{-1}$ 
and have an apparently normal distribution centered on the median value of -40.5 km s$^{-1}$. 
The rms deviation for the sample was 4 km s$^{-1}$.  Assuming that the distribution          
is Gaussian, then the uncertainty in the median value is 4/$\sqrt{29}$,  or 0.7 km s$^{-1}$.  
Correcting for the velocity of HD 72660 then yields a velocity of -37.6 $\pm$ 0.7 km s$^{-1}$ 
for Y Car B.  May 5, 2002 (JD 2452399.5) is the date
midway between the two days on which the STIS observations were obtained.

To test the effects of noise on the calculated velocities, we cross-correlated                
a Y Car spectrum which was made up of the 5 exposures taken on 03 May with a spectrum         
composed of the 3 exposures made on 07 May. To improve the S/N, both spectra were smoothed    
with a 20 point boxcar. In this case, only 6 of the orders generated reliable velocity         
estimates, giving an average velocity difference of V(May 3) - V(May 7) =
 7.4 $\pm$ 0.6 km s$^{-1}$. While         
the uncertainty is in agreement with expectations, the velocity difference is much larger     
than can be easily explained by instrumental effects.





\section{Results}


The orbit of the Cepheid Y Car A was redetermined using new velocities
obtained by Kienzle (2000, private communication).
The STIS observation was taken at phase 0.10 in this orbit.    
The Cepheid orbital plus systemic velocity at this phase is 
-24.4 km sec$^{-1}$, or -11.9 km sec$^{-1}$ with respect to 
the systemic velocity.  If the companion, Y Car B, had a mass 
equal to the Cepheid, its measured velocity (orbital plus 
systemic) at this phase would be -0.6 km sec$^{-1}$.  A companion less
massive than the Cepheid would have an even more positive velocity.

Fig. 1 shows that this 
is clearly not what was measured for Y Car B.  The simplest 
explanation is that the companion is itself a short period binary, 
and hence may have large velocities from the short period orbit
added to the velocities from the orbit with the Cepheid.  
Unfortunately, it would require a large number of observations 
in the ultraviolet to determine the short period orbit.
A binary companion would also be consistent with 
one aspect of the data which we originally found 
perplexing, the fact that there is a 7  km sec$^{-1}$ difference 
between velocities taken four days apart, but with a small
(0.6 km sec$^{-1}$) uncertainty in this difference.  Here we are 
presumably seeing motion in the short period orbit.


We note the correlation of the segments of the exposure taken on 
different days demonstrates  that we were able to 
measure a velocity with an interal accuracy of approximately
0.5 km sec$^{-1}$ (for each velocity) with an exposure 
time about a third of our original estimate, a useful 
benchmark for future observations.  

\section{Discussion}


\subsection{Y Car}

 Unfortunately,   the discovery that Y Car is a member of a triple system means that                            
the quite plausible mass estimate obtained by Bohm-Vitense, et al.                            
     (1997), using velocities obtained from HST GHRS (Goddard High Resolution                      
          Spectrograph) spectra was apparently entirely fortuitous.  The                                
  fact that the velocity amplitude looked reasonable must have been                             
       simply due to coincidence - the Y Car B binary phase must have been                           
    such that it did not greatly alter the motion expected due to the                             
     orbit of the original companion (i.e. Y Car B itself) around the Cepheid.                     
     The large measurement uncertainties in the GHRS data would also help                          
     to mask any modest effects of the third star on the velocity of Y Car B.                      
     For example, one spectrum was very weak, while the other was made through                     
      the large science aperture before  the COSTAR (Corrective                     
           Optics Space Telescope Axial Replacement) was in place to correct                             
     spherical                                                                                          
          aberation, and centering of the target in the large aperture was
          more uncertain than in later spectra. The resultant velocity
          measurement uncertainty was estimated to be 7 km/sec, much larger
          than for other stars measured by the same authors.
For reference, the previous GHRS observations were made at orbital 
phases 0.728 and 0.025. We based the mass estimate on the velocity difference,
between the two
rather than the individual velocities to control against differences due 
to differences between Y Car B and the template spectrum.
However, for comparison with the STIS result here,  
for phase 0.728, the velocity was -35.2 or 
-33.4 km sec$^{-1}$ 
depending on whether the comparison star was HD 72660 or $\alpha$ Lyr.
For phase 0.025, the velocity was -8.0 or -2.0 sec$^{-1}$ 
for the same comparison stars. Figure 1 shows that those are 
reasonable velocities for the companion at the respective orbital
phases, although we based our mass estimate on the orbital velocity 
amplitude rather than the absolute velocities.

For the present STIS spectrum, we  
can compare the velocity difference over 4 days with the 
possible periods in short period orbits.  The observed 
velocity implies an orbital velocity semi-amplitude of 
at least 50 km sec$^{-1}$ for the companion (Fig. 1).  The 7 km sec$^{-1}$
velocity difference over 4 days could result from a period
near 4 days or even a period half this length.  On the 
other hand,  4 days  could be only a fraction
of an orbit.  The simplest guess for an orbit (two linear
segments) provides a crude estimate of the period of 20 
days in this case.  These estimates from the velocity
change are reasonable for the short period, and would be 
compatible with a hierarchal structure for the system.

\subsection{Triple Systems}

Triple systems have been a continuing surprise in the study of 
Cepheid binaries, particularly since ultraviolet spectroscopy 
has become available to observe the secondary in detail. 
Typically, but not always,  a third star in the system is found when the 
spectrum of the second star becomes available, e.g. in 
velocity measures of the secondary.  In systems containing 
a cool star with a hot companion, there are also a number 
of ways of inferring a third star from the energy distribution
(Parsons, 2003)

 In Table 1 we present a summary of available information on 
the multiplicity of Cepheids in binary systems.  
The basis for this list is Table 1A in  Evans (1995), containing
systems for which orbits have been determined. 
In the decade since that list was put together, binary motion
and orbits have been determined for a number of additional 
systems (see Szabados: http://www.konkoly.hu/CEP/intro.html ).  
 Systems with orbits have been added to Table 1.  The source
for the orbits since the Evans summary is listed in Column 6 of 
Table 1, with the period (Column 2) and the eccentricity (Column 3)
taken from that source.  Column 4
indicates the multiplicity of the system, as discussed in the 
reference listed in Column 7.  Column 5 indicates whether 
high resolution spectra have been taken in the ultraviolet to 
measure the velocity of the companion.  H indicates an HST 
spectrum: I indicates an IUE spectrum where there is no HST 
spectrum.  
Table 1 is organized in two sections.  The stars with the most 
complete information and hence the highest probability of determining
the full multiplicity of the system are listed at the top.  Those are 
the stars with both an orbit and also an IUE  low resolution
spectrum of the companion.  Below those are stars with
an orbit (BY Cas, VZ Cyg, and MW Cyg) but which are faint 
and do not have an ultraviolet spectrum of the companion.  
In some cases (e.g. S Sgr and AW Per), the presence of 
a third star was inferred from the combination of a mass function and a 
companion with a mass (from the ultraviolet temperature) too small 
to be compatible with a reasonable Cepheid mass. The stars in this bottom
group have no way to identify a such binary companion.

We provide comments here on two stars in Table 1.

V1334 Cyg: This low amplitude Cepheid is a spectroscopic binary.
It was also has had  repeated visual observations in the past  indicating a
companion with a separation of 0.1", leading to the
double star designation ADS 14859. This orbit is much
larger than that of the spectroscopic binary orbit.
However, a recent HST study has not resolved the system
(Evans, et al., 2005a), so we   still do not understand the
system fully.

FF Aql:  Again, the Cepheid is a member of a spectroscopic 
binary.  A companion 7"  distant does not 
seem to be physically related (Udalski and Evans, 1993).
In addition, the system has repeatedly been resolved by 
speckle interferometry (McAlister, et al. 1989) 
indicating a companion 
with a separation between 7" and the separation of the 
spectroscopic binary companion.  However, the speckle 
measures appear to move with respect to the Cepheid 
much more than can be accounted for by the long 
period orbital motion for that separation.  Apparently
neither ``companion" is physically related to the Cepheid.

An estimate of the frequency of triple systems 
among binary systems can be made from the data in 
Table 1.  We will use the 18 stars which have 
orbits as well as a direct ultraviolet observation 
of the companion
(top 18 in the list), since those are the ones
in which a third star can be detected. 
Eight or 44\% 
of those stars are actually triple systems.  
V1334 Cyg  may be 
a triple, which would raise the percentage to 
50\%.  This is a surprisingly high fraction of systems which 
have more than two stars.  

There are several ways in which 
this summary may under or over estimate the detected percentage:

$\bullet$  Under estimate:   Even the well studied stars in
Table 1 may not have 
a full census of more distant companions since we have made no 
particular effort to determine the relationship to nearby 
resolved stars. 

$\bullet$  Over estimate: There is,
on the other hand, a factor that inflates the 
fraction.  If the secondary is over-massive,
the orbital velocity amplitude of the Cepheid will
be larger, which, of course, makes it more likely 
to be detected and studied.

$\bullet$  Under estimate:  Only 8 of the systems in the 
top list in Table 1 have had velocities of the companions 
measured from high resolution ultraviolet spectra.
Even from that list, we were not able to find a velocity
signal for AW Per B. In addition,  
even a complete high resolution survey could still miss 
face on systems, however such a survey 
would result in a very high 
probability of detection a binary companion.  That is, even our 
18 best studied systems have seriously incomplete information 
for identifying triple systems.

On balance, it seems likely that the fraction of triple systems 
in Table 1 is under estimated.   

There is one further marker from ultraviolet low resolution spectra 
which confirms the identification of triple systems.  
The IUE low resolution spectrum of SU Cyg B shows a strong 
Ga II feature at 1414$\AA$ (Wahlgren and Evans 1998), the 
signature of a HgMn star.
Since the feature is very prominent, 
this provides a way to detect likely triple 
systems from a single low resolution IUE spectrum of the companion.
Because chemically peculiar stars require unusually slow rotation 
to allow for diffusive element settling, they are usually found 
in short period binaries where rotation has been slowed by
synchronization with the orbit.  We note that in the list 
of Evans (1995) of stars with orbital motion 
(but not complete orbits) and IUE observations,
(XX Cen, AX Cir, T Mon, AT Pup, and V465 Mon), one on them 
(T Mon) is also chemically peculiar (Evans, et al. 1999), in this case 
a magnitic star with a spectrum very similar to $\alpha2$ CVn.  

  Tokovinin (2004) has assembled a catalogue of multiple star systems.  
He  discusses the range of period  ratios 
P$_{out}$/P$_{in}$ (outer and inner periods in triple systems)
found for low mass stars within 50 pc. 
For P$_{in}$ $\ge$ 10 days he finds 
P$_{out}$/P$_{in}$ $\le$ 10$^4$.  He discusses the empirical 
description of the upper limit of P$_{out}$/P$_{in}$:

$$ {P_{out}(1-e_{out})3\over P_{in}} \ge 5 $$  

\noindent where e is the eccentricity of the outer orbit.
In Table 2 we list the upper limit on the inner period for
values of P$_{out}$ and e$_{out}$ typical of Cepheid systems.  

The orbital parameters for Y Car predict that the inner period 
should be less than 31 days.  This is an agreement with the 
discussion of the short period earlier in this section. 

One implication of these results demonstrating a large number of short 
period binaries in Cepheid systems is that presumably many of the 
Cepheids themselves were also once short period binaries,  which 
would  have coalesced.  If the result is simply a star 
with the combined mass of the binary pair, it may proceed to become a 
Cepheid with no obvious signs of its former status.  It would, however,
belong to a younger isochrone than the original pair.  

A coalescence scenario has been suggested by Alcock, et al. (1999)
to explain the period distribution of the Cepheids in the LMC.  
Specifically, they find proportionately too many short period 
Cepheids, which they suggest is the result of coalesced lower mass
binaries.  As they point out, this is the scenario for 
``anomalous Cepheids" (which is what some of their short
period stars would be), and also for blue stragglers.
One aspect of this scenario is that the ``blue loops" 
of the He burning phase extend to hotter temperatures
than for single star evolution, and hence reach the 
instability strip.  Blue loops which are too cool have 
also been found, for instance,  in comparing the components of binary 
Cepheids with evolutionary tracks (e.g. Evans, 1995).
Revision of the evolutionary history of the Cepheids 
to include coalescence would produce tracks with 
blue loops extending to hotter temperatures, and 
might resolve this problem.

How does this fraction for supergiants compare with other 
triple fractions?  Tokovinin (2004) 
estimates that the fraction of 
multiple systems with more than 2 stars 
in his catalogue is 20 \%.  However, comparing
the F, G, and K dwarfs in his catalogue and comparable stars in the 
Hipparcos catalogue, he finds the percentage of triple stars 
decreases from 17\% within 10 pc to 2\% within 50 pc.
  He attributes this to serious incompleteness as the distance
  increases, partly 
due to the fact that identifying higher order multiples often 
requires more than one kind of observation.  Evans (1992b) surveyed 
of Cepheids using ultraviolet low resolution spectra from
 IUE.  The fraction of Cepheids with a hot companion from 
the survey (corrected to the fraction which would be detected 
in a reasonably sensitive radial velocity survey) is 34 \%.  
As discussed by Evans (1995), the list of Cepheids with orbits
is quite complete except for the combination of 
periods longer than  10 years and very small
mass ratios.  Since we have found 44 \%\/ of the binaries
have an additional companion, this implies 15 \%\/ of Cepheids
are in high order multiples.  This is comparable to the 
fraction Tokovinin found for the dwarfs within 10 pc.  That is,  
studies of Cepheids with a variety of techniques have produced 
a fraction of multiple systems similar to the best studied
dwarfs.

\section{Summary}

We have obtained a STIS ultraviolet high dispersion Echelle mode spectrum of
the double mode classical Cepheid Y Car. The velocity measured for the hot
companion from this spectrum is not consistent with binary motion, implying
that the companion is itself a short period binary. The fact that the velocity
changed by 7 km sec$^{-1}$ during the 4 days between two segments of the 
observation
confirms this interpretation. We have summarized ``binary" Cepheids which are
in fact members of triple systems. From the list of stars 
without ultraviolet (IUE) spectra, 44\% are triples.
 Possible sources of bias in this fraction are discussed, 
 and we conclude that the
fraction is underestimated.








\acknowledgments

It is a pleasure to thank Bruce Woodgate, Ted Gull, Jeff Valenti,  
Tom Ayres and Don Lindler
for useful conversations about the accuracy of STIS velocities.
An anonymous referee made several comments that helped the 
presentation of the paper.
This research has made use of the SIMBAD database,
operated at CDS, Strasbourg, France.
Financial assistance was provided ST ScI grant  HST-GO-09146.01-A 
to the Smithsonian Astrophysical Observatory and
HST-GO-09146.01-B to Goddard Space Flight Center, and  
from the Chandra X-ray  Center NASA Contract NAS8-39073.

\clearpage










\clearpage

\begin{deluxetable}{lrccccl}
\tabletypesize{\scriptsize}
\tablecaption{Multiplicity 
\label{tbl-1}}
\tablewidth{0pt}
\tablehead{
\colhead{Star} & \colhead{P} & \colhead{Ecc}   & \colhead{Triple}   & \colhead{High} & 
\colhead{Orbit} &  
\colhead{Multip.}  \\
\colhead{} & \colhead{d} & \colhead{}   & \colhead{}   &  \colhead{Res} & 
\colhead{Source} &  
\colhead{Source}  \\

}
\startdata


U Aql & 1856 & 0.16 & y & H & 1 & Evans, et al. 2005b \\ 
FF Aql & 1430 & 0.09 & & & 1 & Evans et al. 1990 \\ 
RX Cam & 1113 & 0.46 & & & 3 & \\ 
Y Car & 993 & 0.46 & y & H & 1 & this paper \\ 
YZ Car & 657 & 0.14 & & & 5 & \\ 
DL Cas & 684 & 0.35 & & & 1 & \\ 
AX Cir & 6532 & 0.19 & & & 5 & \\ 
SU Cyg & 549 & 0.34 & y & I & 1 & Evans and Bolton, 1990 \\ 
V1334 Cyg & 1937 & 0.20 & ? & I & 2 & Evans 2000 \\ 
Z Lac & 381 & 0.01 & &  &  1 & \\ 
S Mus & 505 & 0.08 & & H &  1 & \\ 
AW Per & 13100 & 0.55 & y & I &  1 & Evans, et al 2000 \\ 
S Sge & 676 & 0.23 & y &  & 1 & Evans et al. 1993 \\ 
W Sgr & 1780 & 0.52 & y & &  1 & Massa and Evans, 2005 \\ 
V 350 Sgr & 1108 & 0.27 & & H & 1 & \\ 
V636 Sco & 1318 & 0.26 & y & H &  1 & B\"ohm-Vitense, et al, 1998 \\ 
$\alpha$ UMi & 10969 & 0.66 & y & &  1 & Kamper 1996 \\ 
U Vul & 2510 & 0.58 & &  & 3,4 & \\ & & & & \\
                                                                                
No IUE & & & & \\                                                                                 

BY Cas  & 563 & 0.22 & & &  6 & \\  
VZ Cyg & 725 & 0.05 & & &  6 & \\  
MW Cyg & 441 & 0.04: & & &  3,6 & \\  

& & & & \\  


 \enddata

References for orbit: 1. Evans, 1995; 2. Evans, 2000; 3. Imbert, 1996
4. Gorynya, 2004; 5. Petterson, et al. 2004; 6. Gorynya, et al. 1995

\end{deluxetable}

\begin{deluxetable}{rrrr}
\tabletypesize{\scriptsize}
\tablecaption{P$_{in}$: Upper Limits (Days)
\label{tbl-2}}
\tablewidth{0pt}
\tablehead{
\colhead{P$_{out}$} & \colhead{e$_{out}$} & \colhead{e$_{out}$}   & \colhead{e$_{out}$} \\
\colhead{d} & \colhead{0.0} & \colhead{0.3} 
  & \colhead{0.7} \\

}
\startdata


365 & 73 & 25 & 2.0 \\
1000 & 200 & 69 & 5.4 \\
3000 & 600 & 206 & 16.2 \\

 \enddata

\end{deluxetable}




\clearpage

\begin{figure} 
\epsscale{0.85}  
\plotone{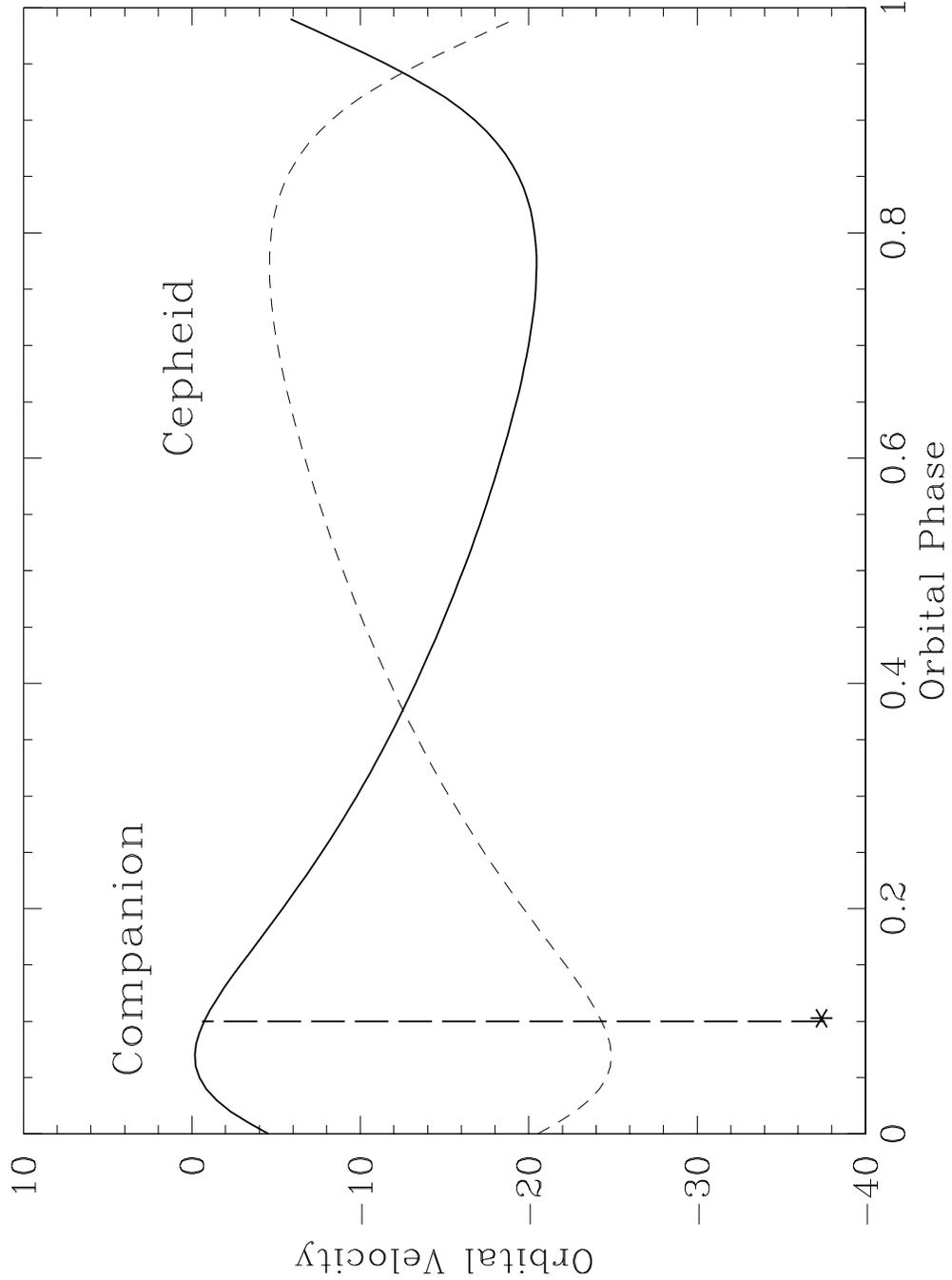}
\figcaption[ycarorb.post.ps]{Orbital velocities. 
The orbital velocity of the Cepheid (dashed
line) is from the ground-based orbit.  The predicted orbital orbital 
velocity of the {\bf companion} ({\bf solid line}), assumes it 
has the same mass as the Cepheid.  The * shows the velocity of the 
companion measured from the STIS spectrum, which is clearly not the 
predicted velocity, as indicated by the vertical dashed line.   
   \label{fig1}}  
                          
\end{figure}


\end{document}